\documentclass[12pt]{article}
\usepackage{amsfonts}
\usepackage{amssymb}
\usepackage{graphicx}
\usepackage{amsmath,graphicx,amsfonts}
\usepackage{epsfig}
\usepackage{dcolumn}

\def\be{\begin{equation}}
\def\ee{\end{equation}}
\begin{document}

%\begin{document}
\title{Darkon fluid - a model for the dark sector of the Universe?\footnote{Updated version of the talk presented by P. Stichel at the XXIX Max Born Symposium, Wroclaw 28-30 June 2011.}}
%\and author{P.C. Stichel$^{1)}$  and W.J.
%Zakrzewski$^{2)}$
%\\
%\siz
%%$^{1)}$An der Krebskuhle 21, D-33619 Bielefeld, Germany \\ \siz
%e-mail:peter@physik.uni-bielefeld.de}
%\author{and W.J.
%Zakrzewski$^{2)}$
%\\ \\ \siz
%$^{2)}$Department of Mathematical Sciences, University of Durham, \\
%\siz Durham DH1 3LE, UK \\ \siz
% e-mail: W.J.Zakrzewski@durham.ac.uk
% }

\author{P.C. Stichel\\%\footnote{peter@physik.uni-bielefeld.de}}
An der Krebskuhle 21, D-33619 Bielefeld, Germany
\thanks{email:peter@physik.uni-bielefeld.de }
\and
Wojtek J. Zakrzewski\\%\footnote{W.J.Zakrzewski@durham.ac.uk}}
Department of Mathematical Sciences,\\ University
of Durham, Durham DH1 3LE, U.K.\thanks{email:W.J.Zakrzewski@durham.ac.uk
}}
%\date{}
\maketitle

\begin{abstract}

%     \section{Abstract}
%      \textbf{Abstract}
%      \vspace{0.7cm}
%INTAN's focus of training -  \emph{attitude},  \emph{skills} and  \emph{knowledge}.

We introduce darkons as fluid particles of a Galilean massless self-gravitating fluid. This fluid exhibits anisotropic scaling with $z=\frac{5}{3}$. The minimal gravitational coupling dynamically generates a gravitational mass density of either sign. Hence such fluid may serve as a model for the dark sector of the Universe.
Its cosmological solutions give a deceleration phase for the early Universe and an acceleration phase for the late Universe. Will the steady flow solutions lead to a confining potential and so a possible model for halos? 
\end{abstract}

      \section{Introduction}
%\textbf{Introductory remarks}

Understanding the dark sector of the Universe is one of the greatest challenges of today's theoretical physics. Is it possible to describe the corresponding astrophysical observations by means of the known concepts
or do we need some kind of `new' physics? 

Dark matter has been introduced already in the thirties of the last century as the missing mass within galaxies in order to understand the galaxies' rotation curves and the motion of individual stars (see [1] for the history of dark matter). But the present model of cold dark matter (CDM) (see the very recent review [2]), although very successful in explaining the large scale experimental data, suffers from two insufficiencies:

1) None of the proposed constituents (particles) has been observed [2],[3].

2) Rotation curves of dark matter dominated galaxies behave in their inner part in sharp contrast to the CDM-based simulations (known as the core-cusp problem [4]). 

Astrophysical observations of the type Ia supernovae and other data (see [5],[6] for recent reviews) suggest that the Universe is undergoing an accelerated expansion. This conclusion was drawn by interpreting the data in the framework of the cosmological Friedmann-Lemaitre-Robertson-Walker (FLRW) model, which has been derived from General Relativity (GR) by assuming that the Universe is homogeneous and isotropic on the largest scales (cf. [7]). Within this framework the cosmic acceleration is attributed to the so-called dark energy (DE) which is the source of repulsive gravitation due to its negative pressure. The simplest DE-model 
uses a positive cosmological constant $\Lambda$. This concept, unified with the CDM-model, leads to the standard $\Lambda$CDM-cosmological model 
which is `favoured by a number of observations' [5]. 
However, when interpreted as the energy density of the vacuum, the experimental value of $\Lambda$ turns out to be a factor of about $10^{121}$ too small ([8],[5]). On the other hand, if we assume that DE has a dynamic origin (corresponding to a time dependent $\Lambda$) one has to change the matter part of Einstein's equations by introducing {\it e.g.} a scalar field which produces the required negative pressure (cf. [5],[6]). Such models, however, have less predictive power as one can always construct a potential that gives rise to a given cosmological  evolution.
Modifying the geometric part of Einstein's equation is another possibility (cf [5],[6]). Such models, however, suffer from having to rely on an arbitrary function which is not derivable from any known fundamental assumption. 

In summary, so far we have no model which has been derived from fundamental (conventional) physics which can explain the late-time cosmic acceleration. Of course, by making this statement we assume that the accelerating expansion of the Universe is a real effect and not an apparent or transient effect due to the averaging over large scale inhomogeneities in the Universe  (cf [9],[10]).

Very recently, the present authors, presented a new model of the dark sector of the Universe, which contains no new parameters at the microscopic level and which is based only on some well known physical 
principles (Galilei symmetry and Einstein's equivalence principle) [11], [12].
 Our model is nonrelativistic but this in accordance with the well-known fact that the FLRW-model can be derived from Newtonian gravity (by heuristic arguments one can even include pressure [13]).

 We note that if we want to consider some new nonrelativistic particles as the source of the accelerating expansion of the Universe these particles must necessarily be massless (massive particles always lead to attractive gravitation). Nonrelativistic massless particles can exist as a dynamical realization of the unextended Galilei algebra. Furthermore, they live necessarily in an enlarged phase space [14] and possess a modified relation between energy and momentum (or velocity).  The gravitational coupling of these `exotic particles' [11] was introduced in the simplest way that satisfies Einstein's equivalence relation. This coupling, for a self-gravitating fluid, leads to a dynamically generated active gravitational mass density of either sign which then becomes a source of the corresponding gravitational field. This fact leads to the possibility of using our model as a basis of further models which can then constructed to explain the observed accelerating expansion of the Universe and also of the CDM-attributed halos which confine the galaxies.

The equivalence principle, {\it i.e.} the local equivalence between gravitation and acceleration is normally formulated mathematically as an invariance w.r.t. arbitrary time-dependent translations (gauge transformations). By fixing this gauge we can reduce the phase space and remove the unphysical variables. The Lagrangian of the resultant model does not possess a free particle limit and hence the particles it describes, darkons, exist only as fluid particles of a self-gravitating fluid. The darkon fluid presents a dynamical realisation of the zero-mass Galilean algebra extended by anisotropic dilation symmetry with the dynamical exponent $z=\frac{5}{3}$. 

The paper is organised as follows. In Section 2 we introduce our Galilean massless particles and discuss their coupling to gravity.
In Section 3 we describe the generalization of this particle picture to a self-gravitating fluid. In Section 4 we consider the cosmological solutions and in Section 5 we state the main open problems. We conclude in Section 6 with some final remarks.

\section{Galilean massless particles and their coupling to gravity.}
%\textbf{Galilean massless particles}

Our main task involves finding the minimal dynamical realization of the unextended ($m=0$) Galilean algebra in $d=3$.
To do this we consider the first order  Lagrangian in a 12-dim phase space
\be
 L_0\,=\,p_i\dot x_i\,+\,q_i\dot y_i\,-p_iy_i\ee
leading to the  equations of motion (EOM)
\be \dot x_i\,=\,y_i,\quad \dot p_i\,=\,0,\quad \dot q_i\,=\,-p_i,\quad \hbox{and} \quad \dot y_i\,=\,0.\ee
    
Here $x_i$, resp. $y_i$ are the components of the position, resp. velocity, of a particle and $p_i$, resp. $q_i$, are the corresponding canonical momenta. They satisfy, in the framework of the equivalent Hamiltonian dynamics, the Poisson-bracket algebra (PBs):
\be [x_i,\,p_j]\,=\,\delta_{ij}\quad \hbox{and}\quad [y_i,\,q_j]\,=\,\delta_{ij}\ee
with the Hamiltonian $H$ given by

\be H\,=\,p_iy_i.\ee

We note that 
\begin{itemize}
\item The relation between energy $H$ and the momenta (velocities) is nonstandard,
\item Momenta $p_i$ and velocities $y_i$ are dynamically independent,
\item The $q_i$ have no counterpart in standard particle mechanics.
\end{itemize}

By Noether's theorem we obtain from (1) 
the boost generator $K_i$ 
\be
K_i\,=\,p_it\,+\,q_i,\ee
which, according to (3) has vanishing PBs with the generator of the spatial translations $P_i=p_i$:
\be [K_i,\,P_j]\,=\,0.\ee

So (6) clearly shows that the Lagrangian (1) describes a dynamical realization of the unextended ($m=0$) Galilei algebra $G_0$. To prove that this realization of $G_0$ is the minimal one we refer to [14].
We note that $G_0$ may be enlarged by dilations $D$ with arbitrary dynamical exponent $z$.  

$D$ is given by [14]:

\be
D\,=\,tH\,-\,\frac{1}{z}x_ip_i\,+\,\left(1-\frac{1}{z}\right) y_iq_i.\ee

To couple our massless particles to the gravitational field strength $g_i(\vec x,t)$ we have to be consistent with Einstein's equivalence principle: {\it Locally the gravitational field strength is equivalent to an accelerating frame}. 

So, the EOM 
\be 
\ddot x_i\,=\,\dot y_i\,=\,0
\ee
for a free particle have to be replaced, in the presence of gravity, by
\be
\ddot x_i(t)\,=\,g_i(\vec x(t),t).
\ee

This EOM  can be realised if we add to $L_0$ an  interaction part {{(minimal coupling)}}
\be L_{int}\,=\, -q_ig_i.\ee
Then the EOM \be\ddot q_i=-\dot p_i=0\ee
 gets replaced by
\be
\ddot q_i\,=\,-\dot p_i\,=\,q_k\,\partial_i g_k.\ee

{\bf Note}:
 Einstein's equivalence principle still holds if we add in (10) further 
terms containing spatial derivatives of $g_i$  ie by replacing
\be g_i\,\rightarrow\,g_i\,-\,K_1\,\triangle g_i\,\quad +\hbox{higher}\quad \hbox{order}\quad\hbox{terms}\ee
where $\triangle$ is the Laplacian.

\section{A self-gravitating (darkon) fluid}
%\subsection{Lagrangian formulation}
In the Lagrangian formulation
 we generalise the one-particle phase space coordinates $A_i\in(x_i,p_i,y_i,q_i)$ to the continuum labelled by $\vec \xi\in R^3$ 
(comoving coordinates)
\be A_i(t)\,\rightarrow \,A_i(\vec \xi,t)\ee
The Lagrangian for the selfgravitating fluid then becomes (after elimination of $p_i$ and $y_i$)
\be
L=-n_0\,\int\,d^3\xi\left(\dot q_i(\vec \xi,t)\dot x_i(\vec \xi,t)\,+\,q_i(\vec \xi,t)g_i(\vec x(\vec \xi,t),t)\right)\,+\, L_{\phi}, \ee
where, as usual,

\be L_{\phi}\,=\,-\frac{1}{8\pi G}\,\int\,d^3x\,(g_i(\vec x,t))^2.\ee

$G$ is Newton's gravitational constant and $n_0$ is the (constant) particle density in $\vec \xi$-space.
{ Euler-Lagrange EOM:}

Varying $q_i$ and $x_i$ leads to the continuum generalisation of the previous equations (9) and (12) for $x_i$ and $q_i$.
    
Now we have two possibilities to go further:

In the first case we
take $g_i=-\partial_i\phi$ and vary $\phi$ getting
\be \triangle \phi\,=\,4\pi G\,\partial_k(nq_k)\ee
with 
\be
n(\vec x,t)\,=\,n_0\,J^{-1}\quad\hbox{(particle}\quad \hbox{density)}\ee
and
\be
J\,=\,\det\left(\frac{\partial x_i(\vec \xi,t)}{\partial \xi_k}\right)\quad \hbox{(Jacobian)},\ee
    where $\vec \xi(\vec x,t)$ denotes the function inverse to $\vec x(\vec \xi,t)$.

 The right hand side of the Poisson equation (17) describes a dynamically generated active gravitational mass density which may be of either sign with

+ sign leading to an attractive gravitation

 - sign leading to a repulsive gravitation.

This promotes the self-gravitating fluid to a possible candidate for the dark sector of the Universe.

In this case our EOM are invariant w.r.t. arbitrary time-dependent translations ({ gauge transformations})
\be
 x_i\,\rightarrow \, x_i\,+\, a_i(t),\ee
where $\phi$ transforms as
\be \phi(\vec x,t)\,\rightarrow \,\phi'(\vec x',t)\,=\,\phi(\vec x,t)\,-\,\ddot a_i(t)x_i\ee
and $q_i$ and $n$ remain invariant.

 In the second case we
vary $g_i$ instead of $\phi$ and this leads to a linear relation between $q_i$ and $g_i$
\be
g_i(\vec x,t)\,=\,-4\pi\, G\, n(\vec x,t)\,q_i(\vec \xi(\vec x,t)t).\ee

 Note that the expression for $g_i$ can be rewritten as
\be q_i(\vec \xi,t)\,=\,-\frac{1}{4\pi G}\,\left(\frac{g_i}{n}\right)(\vec x(\vec \xi,t),t)\ee

The expression (23) allows to  eliminate from the EOM the 
unphysical phase space variables $q_i$ in favour of the gravitational field strength $g_i$.

This removes the unphysical degrees of freedom. However, this expression for $q_i$ (a particular solution of the
 Poisson equation) destroys the gauge symmetry {\it ie} it fixes the gauge.

{\bf Remark}:
The gauge symmetry can be restored by considering a curl-free velocity field 
$$\vec u\,=\,\vec \nabla u$$
and treating $u$ as the variable to be varied [12].

Inserting  $q_i$ from (23) into the previous EOM we get two coupled EOM for $\vec x(\vec \xi,t)$ and $\vec g(\vec x(\vec \xi,t),t)$
\be \frac{d^2}{dt^2}\,x_i(\vec \xi,t)\,=\,g_i(\vec x(\vec \xi,t)t)\ee
(initial condition $\vec x(\vec \xi,0)=\vec \xi$) and
\be \frac{d^2}{dt^2}\,\left(\frac{g_i}{n}\right)(\vec x(\vec \xi,t),t)\,=\,\left(\frac{1}{2n}\,\partial_i\,g_k^2\right)(\vec x(\vec\xi,t),t).\ee
The last equation can be integrated once.
To do this we note that
\be\theta_i\,=\,\frac{\partial \dot x_k}{\partial \xi_i}\,\frac{g_k}{n}\,-\,\frac{\partial x_k}{\partial \xi_i}\,\frac{d}{dt}\left(\frac{g_k}{n}\right)\ee
is conserved ({\it ie} $\frac{d}{dt}\theta_i=0$).

This conservation law 
gives the required once-integrated form of (24)

\be\frac{d}{dt}\,\left(\frac{g_i}{n}\right)\,=\,\frac{g_k}{n}\partial_i\,\dot x_k\,-\,\theta_k\,\frac{\partial \xi_k}{\partial x_i},\ee
which is the transport equation for the field strength $\vec g$ (cf. [15] for the standard Newtonian case).

\vskip 0.3cm
Note that
\be
Q\,=\,\int\,d^3\xi\,\theta_k(\vec \xi)\,\alpha_k(\vec \xi)\ee
generates infinitesimal relabeling transformations 
\be\vec \xi\,\rightarrow\,\vec \xi\,+\,\vec\alpha(\vec \xi)\qquad \hbox{with}\quad \vec \nabla_{\xi}\vec \alpha=0.\ee

 Remark:   

The EOM (24), (27) do not contain the gravitational constant $G$. By inserting (23) into (15) we see that the Lagrangian contains $G$ only 
as a common factor. The Lagrangian does not split into free and interacting parts. Thus the fluid particle (called darkons) do not exist
as free particles but are only elements of a self-gravitating fluid. This comes from the reduction of the phase space 
(expression of $q_i$ in terms of $g_i$ by (23)).

For brevity we dispense with the Eulerian formulation of the darkon fluid dynamics. Therefore we will only summarize the resulting space-time symmetries [12],[14]:

The symmetry algebra becomes the expansion-less conformal Galilei algebra with $z=\frac{5}{3}$. For free particles dilation symmetry can have any $z$, the minimal gravitational coupling fixes $z$ to $z=\frac{5}{3}$ (higher-order derivative terms (13) change this).

\section{Cosmological solutions}

We consider the darkon fluid as a model for the dark sector of the Universe. Hence we look for solutions of the fluid EOM satisfying the cosmological principles (the Universe
being isotropic and homogeneous on large scales).

This gives in the Lagrangian formulation
\be
n(\vec x,t)\,=\,n(t),\quad x_i(\vec \xi,t)\,=\,a(t)\xi_i,\ee
 where $a(t)$ is the cosmic scale factor,
   
and 
\be g_i(\vec x,t)\,=\,-4\pi\,n(t)\,g(a(t))\,x_i\ee
Note that the derivative terms in the gravitational coupling would not contribute!.

If we next put all this into the EOM  we obtain:
    
1. From the definition of $n(\vec x,t)$ 
\be 
n(t)\,=\,\frac{n_0}{a^3(t)}
\ee

2. From the  EOM (24) and using (32)

\be
\ddot a(t)\,=\,-\frac{4\pi n_0}{a^2}g(a),
\ee
{\it i.e.} a Friedmann-like equation

3. From the EOM for $g_i$ and the definition of $\theta_i(\vec \xi)$
\be
\dot g(a(t))\,=\,\frac{\beta}{a^2(t)},\quad{\beta=}\hbox{const},
\ee
 where $\beta$ is related to $\theta_i(\xi)$ by
$$\theta_i(\vec \xi)\,=\,4\pi\beta\xi_i.$$

The last two equations, after integration, give us a cubic equation for $g(a)$ [11]
\be
g(g^2\,+\,C_1)\,+\,C_0\left(1-\frac{a_t}{a}\right)\,=\,0,\ee
with $C_i$ integration constants and
\be
a_t\,=\,\frac{3\beta^2}{2\pi n_0C_0}.
\ee

Next we take $C_i>0$ and use $a(t)$ as a measure of time (given that now we have $\dot a(t)>0$ (expanding Universe))

Then we conclude from (35) and (33) that

1. For $a<a_t$ we have $g(a)>0$ and so $\ddot a<0$ - {\it ie} the deceleration phase of the early Universe.

2. For $a>a_t$ we have $g(a)<0$ and so $\ddot a>0$ and we are in the accelaration phase of the late Universe.

\eject

{\bf Notes} 
\begin{itemize}
\item $a_t$ defines the point of transition between the phases. All this is consistent with cosmological observations. (cf. [11])

 \item We need $C_i>0$ for these results (this has to be explained by some physical arguments)

  \item If $C_i=0$ we reproduce the scale invariant solution valid at small $t$ (then $a(t)\sim t^{\frac{3}{5}}$).
\end{itemize}

\section{Open problems}
%\textbf{Steady flow solutions}
The main open problems are the following:

\begin{itemize}
\item Steady flow solutions. 
{{Question}}: Can a steady darkon fluid flow be a model for halos?
{\it ie} provide a gravitational potential for the formation of stable structures (galaxies or clusters of galaxies)?

For that we need to find steady solutions of our EOM leading to a potential $\phi$ which confines massive objects. 
Our preliminary results for the spherically symmetric case demonstrate the existence 
of such a confining potential at least for small distances from the centre. But to get a 
quantitative agreement with observations we, probably, have to include spatial derivative terms 
in the gravitational coupling (cf. eqs (10) and (13)).

\item Embedding of the model into a larger one which possesses a general relativisitic limit (as $z=\frac{5}{3}$ our model cannot be a non-relativistic limit of a realtivistic one ($z=1$)).

\end{itemize}

\section{Final remarks}
%\textbf{Final remarks}

%\subsection{Achievements}
%\textbf{Achievements}
Our main results are:
\begin{itemize}
\item Parameter free Lagrangian for a self-gravitating system of Galilean massless particles (darkon fluid)
\item Dynamical generation of an active gravitational mass density of either sign.
\item Explanation of the deceleration phase of the early Universe and of the acceleration phase of the late one.
\end{itemize}

%\subsection{Open problems}
%\textbf{Open problems}
%\begin{itemize}
%\item Determination, from physical arguments, of unknown integrations constants (like $C_i$)
%\item Existence of halos?
%\item Embedding of the model into a larger one which possesses general relativistic limit (as $z=\frac{5}{3}$ our model 
%\textcolor{blue}{{cannot}} be a nonrelativistic limit of a relativistic one ($z=1$)).
%\item ?
%\end{itemize}

\end{document}